\documentclass[10pt,aps,prx,twocolumn,superscriptaddress]{revtex4-2}
\usepackage{amsmath,amssymb}
\usepackage{bm,mathtools}
\usepackage[breaklinks=true,colorlinks,citecolor=blue,linkcolor=blue,urlcolor=blue]{hyperref}
\usepackage[pdftex]{graphicx}
\usepackage{float}
\restylefloat{algorithm}
\usepackage{txfonts}
\usepackage{braket}
\usepackage{algorithm}
\usepackage{algpseudocode}
\usepackage{siunitx}
\usepackage{xcolor}
\usepackage{booktabs}
\usepackage{orcidlink}
\usepackage[section]{placeins}
\usepackage{hyperref}
\usepackage{url}

\usepackage{hyperref}

\algdef{SE}[SUBALG]{Indent}{EndIndent}{}{\algorithmicend\ }%
\algtext*{Indent}
\algtext*{EndIndent}

\begin{document}
\title{Quenched Quantum Feature Maps}

\author{Anton Simen$^{\orcidlink{0000-0001-8863-4806}}$}
\email{anton.simen@kipu-quantum.com}
\affiliation{Kipu Quantum GmbH, Greifswalderstrasse 212, 10405 Berlin, Germany}
\affiliation{Department of Physical Chemistry, University of the Basque Country UPV/EHU, Apartado 644, 48080 Bilbao, Spain}

\author{Carlos Flores-Garrigos}
\affiliation{Kipu Quantum GmbH, Greifswalderstrasse 212, 10405 Berlin, Germany}
\affiliation{IDAL, Electronic Engineering Department, ETSE-UV, University of Valencia, Avgda. Universitat s/n, 46100 Burjassot, Valencia, Spain}

\author{Murilo Henrique De Oliveira$^{\orcidlink{0000-0001-9777-8342}}$}
\affiliation{Kipu Quantum GmbH, Greifswalderstrasse 212, 10405 Berlin, Germany}

\author{Gabriel Dario Alvarado Barrios$^{\orcidlink{0000-0002-8684-4209}}$}
\affiliation{Kipu Quantum GmbH, Greifswalderstrasse 212, 10405 Berlin, Germany}

\author{Juan F. R. Hernández$^{\orcidlink{0009-0007-6933-186X}}$}
\affiliation{Kipu Quantum GmbH, Greifswalderstrasse 212, 10405 Berlin, Germany}

\author{Qi Zhang$^{\orcidlink{0000-0001-6223-5516}}$}
\affiliation{Kipu Quantum GmbH, Greifswalderstrasse 212, 10405 Berlin, Germany}

\author{Alejandro Gomez Cadavid$^{\orcidlink{0000-0001-8863-4806}}$}
\affiliation{Kipu Quantum GmbH, Greifswalderstrasse 212, 10405 Berlin, Germany}
\affiliation{Department of Physical Chemistry, University of the Basque Country UPV/EHU, Apartado 644, 48080 Bilbao, Spain}

\author{Yolanda Vives-Gilabert$^{\orcidlink{0000-0002-3744-5893}}$}
\affiliation{IDAL, Electronic Engineering Department, ETSE-UV, University of Valencia, Avgda. Universitat s/n, 46100 Burjassot, Valencia, Spain}

\author{José D. Martín-Guerrero$^{\orcidlink{0000-0001-9378-0285}}$}
\affiliation{IDAL, Electronic Engineering Department, ETSE-UV, University of Valencia, Avgda. Universitat s/n, 46100 Burjassot, Valencia, Spain}
\affiliation{Valencian Graduate School and Research Network of Artificial Intelligence (ValgrAI), Valencia, Spain}

\author{Enrique Solano$^{\orcidlink{0000-0002-8602-1181}}$}
\affiliation{Kipu Quantum GmbH, Greifswalderstrasse 212, 10405 Berlin, Germany}

\author{Narendra N. Hegade$^{\orcidlink{0000-0002-9673-2833}}$}
\affiliation{Kipu Quantum GmbH, Greifswalderstrasse 212, 10405 Berlin, Germany}

\author{Archismita Dalal$^{\orcidlink{0000-0003-0638-8328}}$}
\email{archismita.dalal@kipu-quantum.com}
\affiliation{Kipu Quantum GmbH, Greifswalderstrasse 212, 10405 Berlin, Germany}

\date{\today}

\begin{abstract}
We propose a quantum feature mapping technique that leverages the quench dynamics of a quantum spin glass to extract complex data patterns at the quantum-advantage level for academic and industrial applications.
We demonstrate that encoding a dataset information into disordered quantum many-body spin-glass problems, followed by a nonadiabatic evolution and feature extraction via measurements of expectation values, significantly enhances machine learning~(ML) models.
By analyzing the performance of our protocol over a range of evolution times, we empirically show that ML models benefit most from feature representations obtained in the fast coherent regime of a quantum annealer, particularly near the critical point of the quantum dynamics. 
We demonstrate the generalization of our technique by benchmarking on multiple high-dimensional datasets, involving over a hundred features, in applications including drug discovery and medical diagnostics.
Moreover, we compare against a comprehensive suite of state-of-the-art classical ML models and show that our quantum feature maps can enhance the performance metrics of the baseline classical models up to 210\%.
Our work presents the first quantum ML demonstrations at the quantum-advantage level, bridging the gap between quantum supremacy and useful real-world academic and industrial applications.
\end{abstract}

\maketitle
\section{Introduction}
Quantum computing has the potential to accelerate machine learning~(ML) algorithms beyond the classical limits. In particular, researchers have identified provable quantum speedups for certain fundamental ML subroutines, ranging from solving linear systems of equations~\cite{Harrow2009} to performing kernel-based classification~\cite{Rebentrost2014}. Beyond the benefits of asymptotic speed-ups, quantum processors can also generate high-dimensional quantum states or probability distributions that are intractable to simulate classically~\cite{Arute2019,Zhong2020}. 
This capability inspires quantum feature mapping approaches~\cite{Havlicek2019, schuld2019quantum}, which embed classical data in entangled quantum states to enrich ML models with features that are irreproducible by classical computers. Despite these promises, state-of-the-art quantum ML algorithms have not yet demonstrated a practical performance advantage on current noisy quantum hardware~\cite{Schuld2022}. However, recent years have seen multiple demonstrations of quantum computational supremacy in specialized tasks on diverse platforms,
demonstrating that quantum hardware can outperform classical supercomputers in carefully chosen benchmarks~\cite{Arute2019, Zhong2020, king2025beyond}. 

One practical paradigm to go beyond proofs of concept is adiabatic quantum computing~\cite{farhi2000quantumcomputationadiabaticevolution, RevModPhys.90.015002}, where dedicated hardware, such as quantum annealers, already features a few thousand qubits. This paradigm aims to prepare the solution to a problem as the ground state of a Hamiltonian. It relies on the adiabatic theorem, which requires slowly evolving a quantum system from a simple Hamiltonian to the problem Hamiltonian. However, this is hard to satisfy in practice, where finite-time evolutions induce transitions to excited states. Finite-time effects, beyond errors in the desired computation, define the natural uncontrolled evolution of the quantum system, which bears inherent intractable complexity, since describing the transitions requires access to the entire instantaneous eigenspectrum~\cite{Berry_2009}. Recent works on quantum critical dynamics demonstrate that quantum annealers can outperform state-of-the-art tensor network simulations, claiming quantum supremacy in quantum simulation~\cite{king2023quantum, king2025beyond}.

In this work, we push the state-of-the-art in quantum supervised learning algorithms, which primarily employ noisy intermediate-scale quantum hardware.
Existing key methods include variational and evolutionary quantum classifiers with adaptable quantum circuits optimized via classical algorithms~\cite{schuld2020circuit, simen2025evolutionary}, quantum kernel methods embedding data in high-dimensional quantum feature spaces~\cite{Havlicek2019}, and quantum convolutional neural networks that exploit layered quantum circuits for local feature extraction with efficient parameter scaling~\cite{cong2019quantum, simen2024digital}. Thus, current quantum classifiers remain proofs of concept requiring significant hardware advances for large-scale industrial problems \cite{schuld2019quantum}.
Here, we bridge the developments in quantum ML and quantum supremacy by transforming the result of a beyond-classical simulation on D-Wave quantum annealing processors~\cite{king2025beyond, king2022coherent, king20235000} into an empirical quantum advantage for an ML task. We leverage quantum feature mapping to embed classically intractable data representations into an ML pipeline. With this, we demonstrate that quantum-enhanced models can outperform purely classical methods on real tabular data classification tasks. 

We show how ML can benefit from quantum feature maps created by nonadiabatic transitions and entangled states from a nonparametrized quantum reservoir. Similar works on quantum reservoir computing have been done by leveraging different large quantum systems in ML tasks~\cite{cimini2025large, kornjavca2024large}. In contrast to mapping a feature vector as the ground state of a spin-glass Hamiltonian~\cite{umeano2024ground},
we drive the system to a highly entangled state through a fast, non-adiabatic evolution. By doing so, one allows for coupling the instantaneous ground state with other instantaneous eigenstates, creating a complex, classically intractable representation of the feature vector. 
We evaluate the performance of quantum feature mapping by applying quantum analog dynamics to challenging datasets, with more than 100 features, from various domains including drug design and clinical research. We conduct a comprehensive benchmarking of our method across multiple ML models and evaluation metrics, consistently observing improvements over all purely classical baselines. In particular, classification performance peaked at annealing times around $20–30$ ns, a regime closer to the critical point, where entanglement is maximized.

The paper is organized as follows. In \S\ref{sec:aqfm}, we explain our analog quantum feature mapping technique and its implementation. Next, we introduce the benchmarking datasets and machine learning models in \S\ref{sec:bench}. We demonstrate the usefulness of our technique using the D-Wave quantum annealer and present our results in \S\ref{sec:results}. Finally, in \S\ref{sec:conclude} we conclude with an outlook of our technique and its applications.

\section{Analog quantum feature mapping}
\label{sec:aqfm}
\begin{figure*}[t]
    \centering
    \includegraphics[width=.9\linewidth]{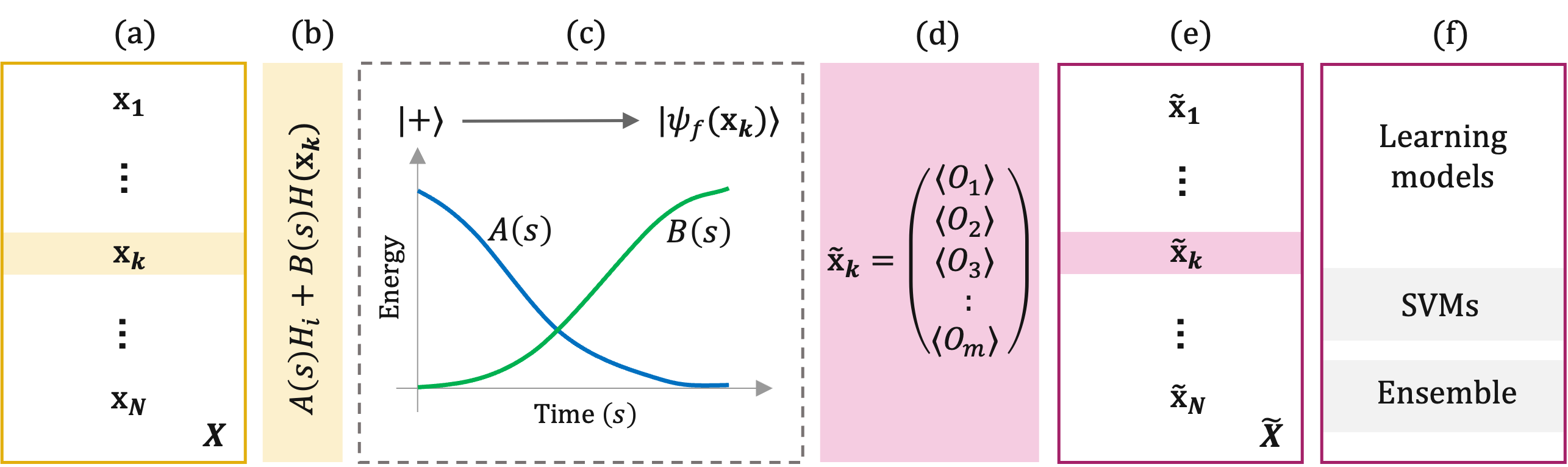}
    \caption{A schematics of analog quantum feature mapping for ML. 
    In the step of quantum feature mapping, (a) the information of a tabular dataset is (b) encoded into the target Hamiltonian, which then (c) performs quantum quench to generate a feature map (d) which (e) composes the quantum-enhanced dataset. 
    This dataset is identified by new feature vectors~$\tilde{\textbf{x}}$.
    This ``quantum" dataset is then used  to (f) train classical ML models such as support vector machines and random forests. 
    }
    \label{fig:aqfm_scheme}
\end{figure*}

An analog quantum feature map~(AQFM) is obtained by performing a quench dynamics of the quantum system encoding the feature vectors of a supervised learning data set.
In particular, we tackle the task of tabular data classification; see the schematics in Fig.~\ref{fig:aqfm_scheme}.
The information of the training data set $\mathcal{X} \in \mathbf{R}^{N\times n}$ is encoded in a set of Ising Hamiltonians.
For each Hamiltonian, $H(\textbf{x})$, we encode the feature vector~$\mathbf{x} \in \mathcal{X}$ into the longitudinal fields, and the classical correlations between the features are encoded as the coupling terms.
We then quench a system following the transverse field Ising model 
\begin{equation}
    \mathcal{H}(s) = -A(s)\sum_{i=1}\sigma^x_i + B(s)H(\textbf{x}),
\end{equation}
using a quantum annealer operating in its coherent regime. The annealing schedules $A(s)$ and $B(s)$, with the normalized time $s = t/\tau$, depend on the quantum hardware. In fast evolutions, when the adiabatic limit is not achieved, the system is driven to a highly entangled state with the presence of nonadiabatic excitations~\cite{king2025beyond}. 
By repeatedly sampling the final state~$|\psi_f(\textbf{x})\rangle$ from the quench dynamics of $H(\textbf{x})$, we can efficiently estimate quantities, $\langle\mathcal{O}\rangle$, such as single-qubit ($\langle \sigma_i^z \rangle $) and multi-qubit ($\langle \sigma_i^z \sigma_j^z \cdots\rangle $) expectations.

We then express the quantum feature representation of a feature $\mathbf{x}$  as
\begin{equation}
    \tilde{\mathbf{x}} = \sum_{i=1}^{n}\langle \psi_f\left(\textbf{x}\right)|\mathcal{O}_i|\psi_f\left(\textbf{x}\right)\rangle \hat{\textbf{e}}_i.
    \label{qfm}
\end{equation}.
For a sufficiently large quantum system with dimensionality beyond 2D, this mapping cannot be efficiently reproduced by a classical method due to the exponential capacity of the final entangled state.
Additionally, expectations of higher-order observables can also be used to enrich the feature set of machine learning models.

In order to apply the analog quantum feature mapping in the context of kernel machines, similarities could be measured through $K(\mathbf{x}_i, \mathbf{x}_j) = \left|\langle \psi_{f}\left(\textbf{x}_i\right)|\psi_{f}\left(\textbf{x}_j\right)\rangle\right|^2$. However, for large systems where the final state is difficult to accurately characterize, it may not find overlaps. Additionally, there are widely used circuit-based approaches for measuring similarities through the evaluation of the unitary $U(\mathbf{x}_i)U^{\dagger}(\mathbf{x}_j)$ \cite{havlivcek2019supervised}, which requires $\mathcal{O}(n^2)$ quantum hardware calls. Thus, we consider the observable measurements obtained in Eq.~\eqref{qfm} to compute similarities as $K(\tilde{\mathbf{x}}_i,\tilde{\mathbf{x}}_j) = \langle\tilde{\mathbf{x}}_i, \tilde{\mathbf{x}}_j\rangle$, which requires $\mathcal{O}(n)$ quantum hardware calls, and then use them to maximize the dual objective function of a support vector machine.
This means that the classical model will learn from the complex patterns inherently encoded into the Gram matrix constructed using the quantum features.

In this work, we use D-Wave's quantum annealers, in their ``fast" regime, to simulate the effect of quantum quench dynamics. As the hardware is not all-to-all connected, we adapt the Ising Hamiltonians to fit the architecture.
We further perform post-processing techniques on quantum feature maps to enhance the performance of the ML models. All such techniques are integrated into a commercial quantum computing product called \textit{Huk}~\cite{HUK}, which is available on the Kipu Quantum Hub~\cite{Quantum_HUB}. Huk is an end-to-end quantum-enhanced machine learning service capable of processing tabular datasets via a quantum computer to generate enriched feature representations, and subsequently use them to train ML models. We use Huk for one of the datasets analyzed in this work.

\section{Benchmarking datasets and models}
\label{sec:bench}
We evaluate binary classification tasks on three publicly available datasets from the UCI Machine Learning Repository. The molecular toxicity dataset~\cite{toxicity_728} contains chemical compound descriptors for predicting toxic activity. The myocardial infarction complications dataset~\cite{myocardial_infarction} includes patient records for predicting post-infarction complications. The drug-induced autoimmunity~(DIA) dataset~\cite{DIA_dataset} contains clinical and chemical data for predicting autoimmune reactions to drugs.
As a demonstration of our Huk service, we employ this service to the DIA dataset, whereas we use AQFM for the other two datasets.

We perform exploratory data analysis to assess structure and quality, identify missing values, and prepare features for model training. In particular, missing values are imputed using the median of each feature to mitigate outlier effects while preserving central tendency. Afterwards, all features are normalized with a standard scaler to achieve zero mean and unit variance, improving comparability across variables.
For the molecular toxicity dataset, we select the top 200 from 1203 features using mutual information with the target variable. For the myocardial infarction dataset, we choose three separate tasks, namely atrial fibrillation, chronic heart failure, and myocardial infarction relapse, among the 11 different binary classification tasks defined on a shared feature set. The three target variables are selected due to their greater representation of the minority class ($>9\%$). For the DIA dataset, variables with zero variance and those with mutual information lower than $0.005$ are removed, creating a dataset with 114 features.

We use four classical machine learning algorithms across all experiments, namely, support vector machine~(SVM), random forest~(RF), gradient boosting~(GB), and extreme gradient boosting (XGBoost). Each model is trained both on the original features and on quantum-enhanced features generated via quantum quench dynamics~(\S\ref{sec:aqfm}), where classical inputs are embedded into quantum states and evolved under a disordered Hamiltonian. Observables from this evolution form the quantum features. 
Models trained on these features are denoted with the “AQFM-” prefix, while those using our advanced classification service Huk are denoted with the “HUK-” prefix. 
This setup ensures that comparisons are made between identical model architectures, isolating the effect of the feature representation.
For the molecular toxicity classification, we also evaluate performance at different annealing times to assess the impact of coherent quantum dynamics.

\section{Results and Discussion}
\label{sec:results}

\begin{figure}[b!]
    \centering
    \includegraphics[width=.9\linewidth]{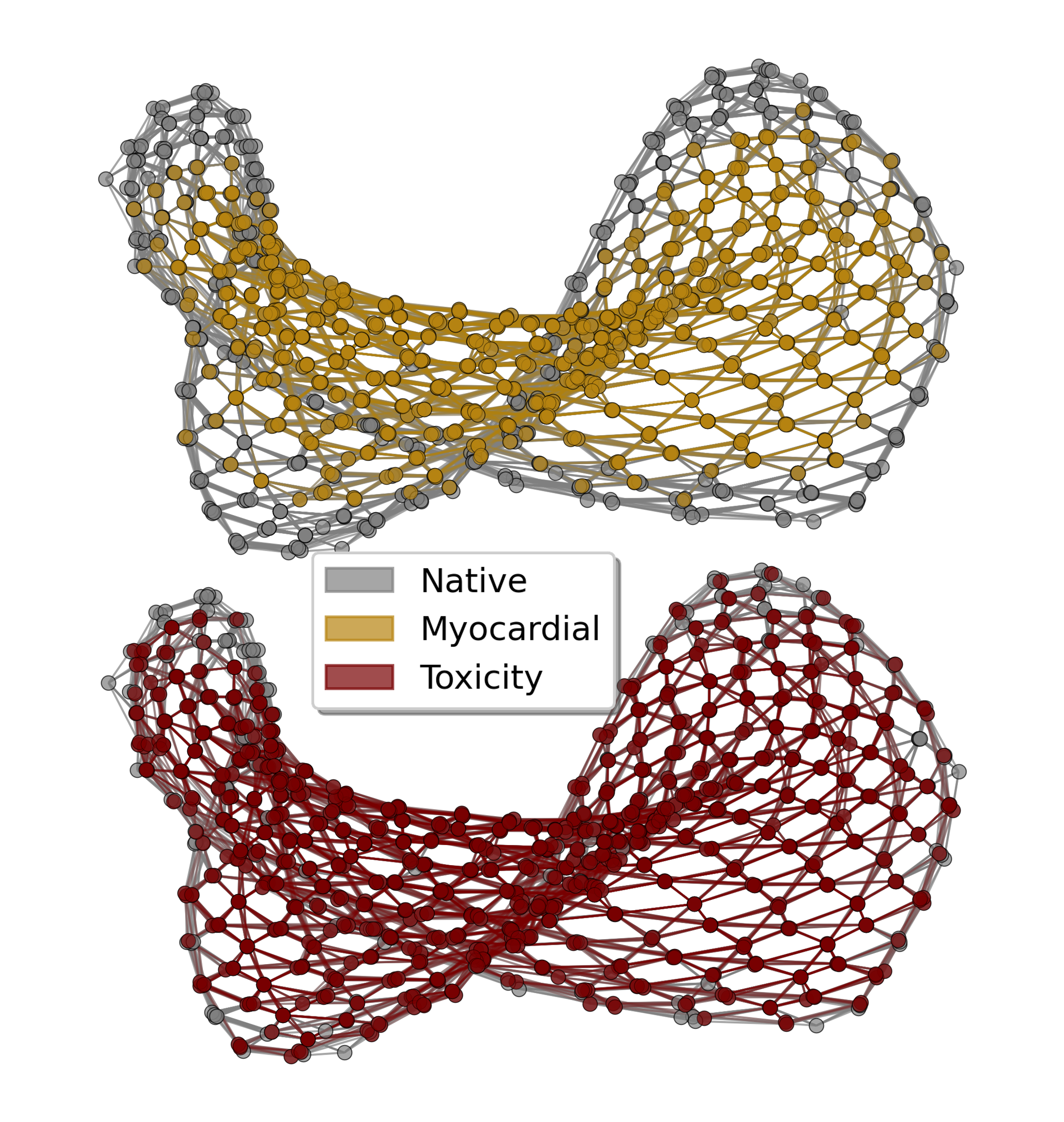}
    \caption{Logical problems for both molecular toxicity and myocardial infarction complications datasets embbeded into Advantage2-prototype2.6.
    }
    \label{fig:embedding}
\end{figure}
This section presents the performance evaluation of machine learning models trained on both standard and quantum-enhanced features. Among the three benchmarking datasets, namely molecular toxicity, myocardial infarction complications and drug induced autoimmunity datasets, with 200, 111 and 114 required qubits, respectively, we evaluate the first two with our AQFM algorithm and the final one using the Huk service. In Fig.~\ref{fig:embedding}, we present embeddings of the logical problems into physical qubits of D-Wave Advantage2-prototype2.6. 
For the first dataset, we also investigate the dependence of model performance on anneal time to study the role of fast coherent dynamics in quantum feature mapping. The section is organized to first outline the evaluation methodology, then analyze performance variation with anneal time, and finally present the results for each dataset.

\begin{figure}[t!]
    \centering
    \includegraphics[width=\linewidth]{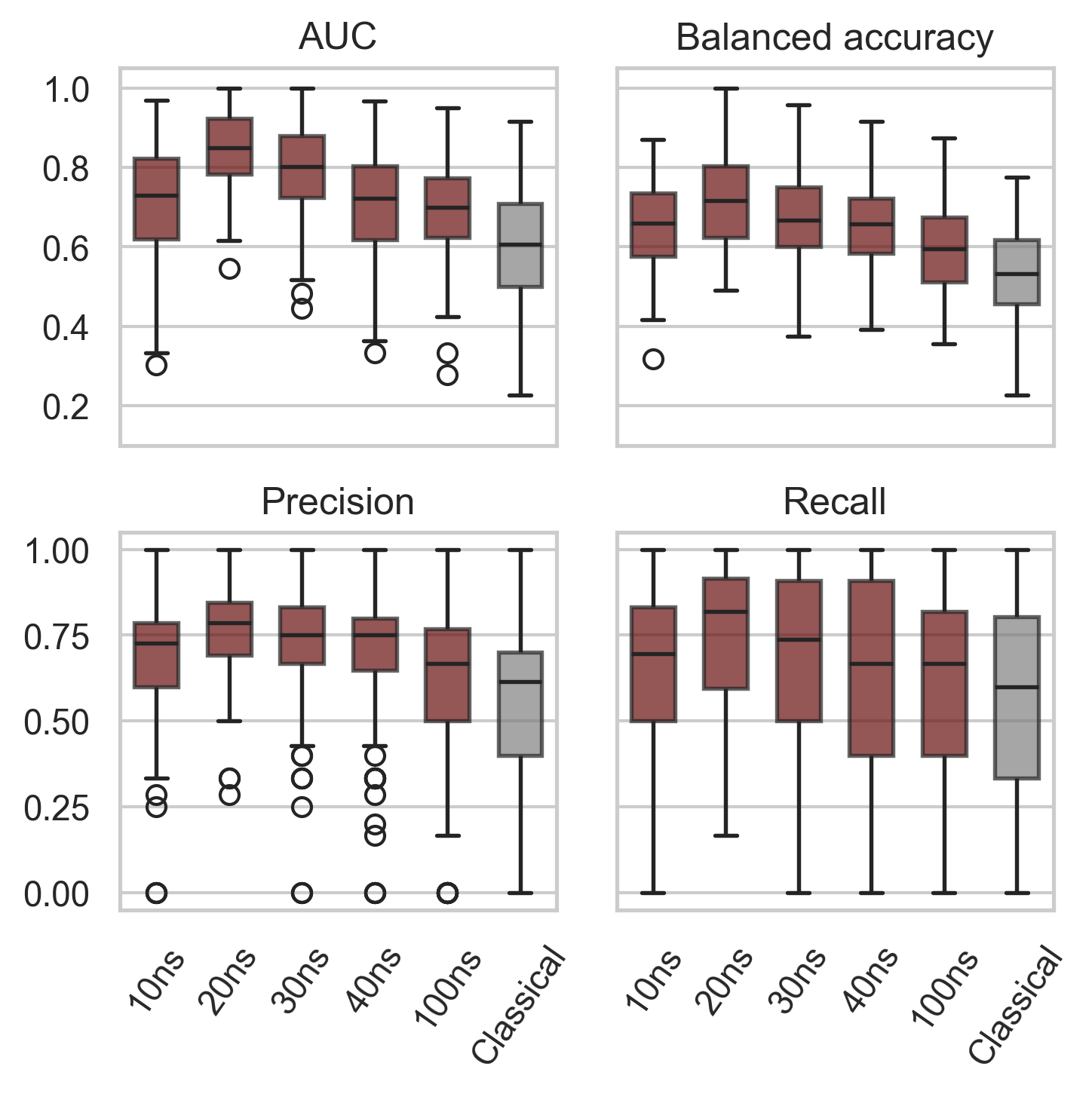}
    \caption{Cross-validation performance comparison across multiple metrics using gradient boosting.
Each of the five plots shows performance as a function of annealing time, with statistically estimated metrics along the y-axis. The chosen annealing times span the coherent regime (10--40 ns) and beyond.
    }
    \label{fig:toxicity_res}
\end{figure}

We follow a standardized and reproducible evaluation pipeline. Given the relatively limited number of samples in the datasets, we adopt a rigorous cross-validation strategy to ensure reliable and statistically meaningful performance estimates. After preprocessing, we optimize hyperparameters via grid search and retrain the best configuration using a StratifiedKFold protocol with 10 splits and 5 repetitions, yielding 50 evaluation scores per model. We report the median score for each metric to provide a more stable estimate of generalization performance, mitigating the impact of sample variability and helping to prevent overfitting. The Huk solver, on the other hand, provides with the performance metrics from a single, given test set, even though it performs cross-validation as well as hyperparameter search on the training dataset. The metrics include accuracy, precision, recall, F1-score, and AUC (Area Under the Curve), providing a balanced view of binary classification performance. These metrics offer a comprehensive basis for comparing the effectiveness of classical and our quantum-enhanced learning pipelines.
\begin{table}[t!]
\centering
\caption{Results for Molecular Toxicity dataset. For each metric, the maximum value across models is highlighted in bold.}
\label{tab:model_toxicity_performance}
\begin{tabular}{lccccc}
\toprule
\textbf{Model} & \textbf{Accuracy} & \textbf{Precision} & \textbf{Recall} & \textbf{F1-Score} & \textbf{AUC}\\
\midrule
GB & 0.54 & 0.40 & 0.33 & 0.53 & 0.62\\
SVM & 0.47 & 0.28 & 0.17 & 0.52 & 0.62\\
RF  & 0.54 & 0.42 & 0.24 & 0.30 & 0.63\\
XGB & 0.56 & 0.37 & 0.30 & 0.51 & 0.56 \\
\hline
AQFM-GB  & \textbf{0.75} & \textbf{0.78} & 0.52 & \textbf{0.77} & \textbf{0.88} \\
AQFM-SVM  & 0.67 & 0.58 & 0.53 & 0.67 & 0.76 \\
AQFM-RF  & 0.60 & 0.68 & 0.22 & 0.57 & 0.81 \\
AQFM-XGB  & 0.69 & 0.67 & \textbf{0.58} & 0.73 & 0.83 \\
\bottomrule
\end{tabular}
\end{table}

We analyze the effect of annealing time on the molecular toxicity dataset by comparing short coherent anneals of 10 ns to 40 ns with longer anneal of 100 ns. Fig.~\ref{fig:toxicity_res} indicates that models using short anneals generally outperform those using longer times across all metrics, including balanced accuracy, precision, recall, F1-score, and AUC. The best results occur at 20 ns and 30 ns, which yield both higher median values and tighter inter-quartile ranges, indicating stronger and more consistent performance. 
Beyond the coherent regime, performance metrics decline and variability increases, likely due to enhanced decoherence effects and thermalization. The classical baseline performs worst across most metrics, confirming that quantum annealing within the coherent regime offers a tangible advantage for this problem instance. 
In the subsequent demonstrations, we use an anneal time of 20~ns.

\begin{table}[t!]
\centering
\caption{Myocardial Infarction Complications dataset results for the prediction of atrial fibrillation episodes.}
\label{tab:model_myocardial_fibr_performance}
\begin{tabular}{lccccc}
\toprule
\textbf{Model} & \textbf{Accuracy} & \textbf{Precision} & \textbf{Recall} & \textbf{F1-Score} & \textbf{AUC}\\
\midrule
GB & 0.62 & 0.22 & 0.41 & 0.58 & 0.70\\
SVM & 0.52 & 0.24 & 0.32 & 0.52 & 0.46 \\
RF      & 0.53 & 0.44 & 0.19 & 0.53 & 0.71 \\
XGBoost  & 0.58 & 0.29 & 0.24 & 0.59 & 0.68 \\
\hline
AQFM-GB  & 0.63 & 0.35 & 0.29 & \textbf{0.63} & \textbf{0.78} \\
AQFM-SVM  & \textbf{0.70} & 0.25 & \textbf{0.65} & 0.60 & 0.77 \\
AQFM-RF  & 0.57 & 0.38 & 0.22 & 0.48 & 0.77 \\
AQFM-XGB  & 0.58 & \textbf{0.46} & 0.25 & 0.57 & 0.76 \\
\bottomrule
\end{tabular}
\end{table}

\begin{table}[t!]
\centering
\caption{Myocardial Infarction Complications dataset results for the diagnosis of chronic heart failure.}
\label{tab:model_myocardial_zsn_performance}
\begin{tabular}{lccccc}
\toprule
\textbf{Model} & \textbf{Accuracy} & \textbf{Precision} & \textbf{Recall} & \textbf{F1-Score} & \textbf{AUC}\\
\midrule
GB & \textbf{0.66} & 0.47 & 0.51 & 0.66 & 0.74 \\
SVM & 0.61 & 0.35 & 0.51 & 0.59 & 0.67 \\
RF  & 0.61 & 0.65 & 0.26 & 0.63 & 0.73 \\
XGBoost  & \textbf{0.66} & 0.47 & 0.48 & 0.65 & 0.72 \\
\hline
AQFM-GB  & \textbf{0.66} & 0.65 & 0.35 & \textbf{0.67} & \textbf{0.75} \\
AQFM-SVM  & 0.65 & 0.39 & \textbf{0.58} & 0.62 & 0.70 \\
AQFM-RF  & 0.58 & \textbf{0.85} & 0.18 & 0.59 & 0.74 \\
AQFM-XGB  & 0.64 & 0.67 & 0.33 & 0.66 & 0.74 \\
\bottomrule
\end{tabular}
\end{table}

\begin{table}[t!]
\centering
\caption{Myocardial Infarction Complications dataset results for the prediction of myocardial infarction relapse.}
\label{tab:model_myocardial_rec_performance}
\begin{tabular}{lccccc}
\toprule
\textbf{Model} & \textbf{Accuracy} & \textbf{Precision} & \textbf{Recall} & \textbf{F1-Score} & \textbf{AUC}\\
\midrule
GB & 0.58 & 0.18 & 0.31 & 0.55 & 0.63 \\
SVM & 0.63 & 0.16 & 0.52 & 0.53 & 0.67 \\
RF & 0.57 & 0.22 & 0.25 & 0.57 & 0.66 \\
XGBoost & 0.52 & 0.20 & 0.14 & 0.52 & 0.64 \\
\hline
AQFM-GB & 0.61 & \textbf{0.25} & 0.32 & \textbf{0.60} & \textbf{0.74} \\
AQFM-SVM & \textbf{0.67} & 0.22 & \textbf{0.63} & 0.56 & 0.71 \\
AQFM-RF & 0.56 & \textbf{0.25} & 0.19 & 0.56 & \textbf{0.74} \\
AQFM-XGB & 0.54 & 0.18 & 0.12 & 0.52 & 0.72 \\
\bottomrule
\end{tabular}
\end{table}

\begin{figure}[t!]
    \centering
    \includegraphics[width=0.95\linewidth]{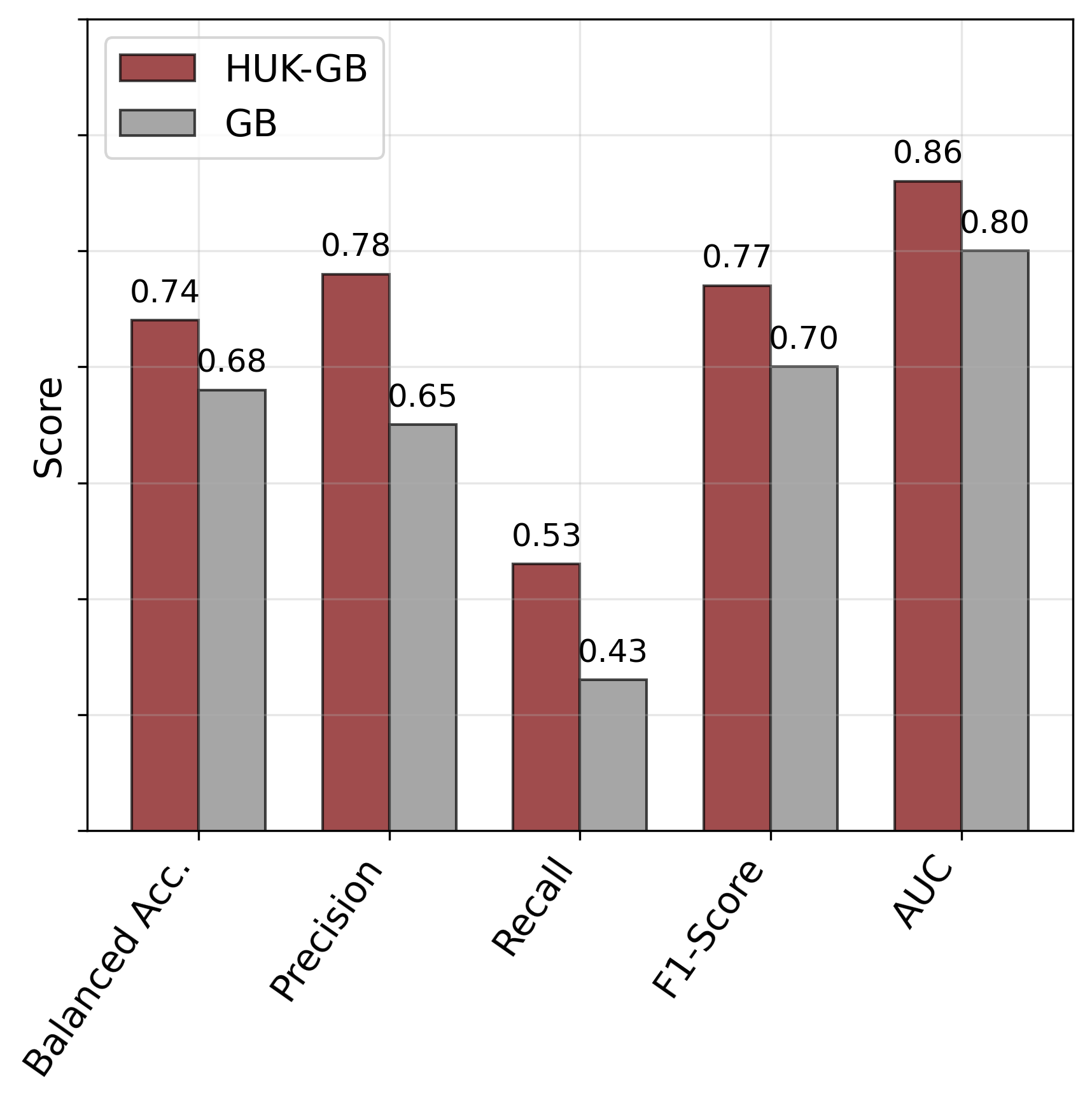}
    \caption{Performance metrics on the DIA  dataset using our Huk service with D-Wave Advantage2\_system1.4 hardware. The bar plot shows test scores across different metrics for gradient boosting with (HUK-GB) and without (GB) the analog quantum feature mapping.}
    \label{fig:dia_results}
\end{figure}

A detailed benchmarking of the three datasets reveals that the AQFM-based models show clear and consistent improvements over their classical counterparts across all evaluation metrics; see Tables~\ref{tab:model_toxicity_performance}, \ref{tab:model_myocardial_fibr_performance}, \ref{tab:model_myocardial_zsn_performance} and \ref{tab:model_myocardial_rec_performance}.
For the molecular toxicity dataset, AQFM-GB yields the best performance in nearly every metric among all eight models, providing a performance increase of at least 40\% over the baseline GB model.
Similar trends are observed for other baselines, with AQFM-enhanced models achieving higher median precision, recall, and F1-scores. These gains indicate that quantum feature mapping via quench dynamics enhances feature quality in ways that benefit multiple classification algorithms.
Across the three myocardial infarction complication tasks, AQFM-based models achieve substantial relative gains over their classical baselines, with improvements exceeding 100\% in recall for atrial fibrillation prediction and up to 58\% in precision for certain configurations. 
Using our advanced classification service for the DIA dataset, the HUK-GB model outperforms the GB baseline across all metrics, with gains of 8.8\% in precision and 7.5\% in AUC, demonstrating consistent enhancement in predictive accuracy; see Fig.~\ref{fig:dia_results}.
All the above results highlight the ability of quantum quench dynamics to extract richer and more informative features, boosting model performance in both molecular and clinical domains. 

Across all cases, the results show substantially higher recall values, indicating that the quantum-enhanced models are better at identifying the minority class. Higher recall means the models are both increasing true positives and reducing false negatives. In medical diagnosis, this metric is critical because correctly identifying more positive cases while missing fewer actual ones can directly impact patient outcomes. In our case, for the prediction of fibrillation episodes, a $24\%$ increase in recall on a test set of 398 samples could mean that dozens more patients are correctly identified as being at risk rather than being overlooked, which in practice could translate into earlier intervention and improved chances of recovery. Similarly, in the molecular toxicity classification task, a $25\%$ increase in recall on a test set of 40 samples means that several additional toxic compounds are correctly flagged, reducing the risk of false negatives and improving safety assessments.

\section{Conclusion}
\label{sec:conclude}
We introduced and experimentally validated an analog quantum feature mapping technique based on the non-adiabatic quantum quench dynamics of quantum spin-glass Hamiltonians. By embedding classical data into disordered many-body systems and extracting features from single-body expectation values, we achieved consistent performance improvements at the quantum-advantage level for prediction of molecular toxicity, diagnosis of myocardial infarction complication, and drug-induced autoimmunity classification. Comparative evaluations against classical baselines showed measurable gains in all metrics tested, with particularly strong improvements in precision, recall, and AUC on several tasks. Systematic benchmarking of quantum annealing times revealed that operating in the coherent regime (10--40 ns) yields the most stable and highest median performance, while longer anneals beyond this regime suffer from degradation due to decoherence and thermalization effects. 

Future research will explore richer quantum feature encodings using many-body correlation measurements, adaptive annealing schedules, and broader problem domains. Such extensions aim to further close the gap between current quantum hardware capabilities and impactful real-world applications, enabling scalable quantum-enhanced machine learning across diverse scientific and industrial problems. 
Additionally, a promising future direction for this work is the implementation of the protocol on digital quantum computers, which offers greater flexibility for encoding classical information and simulating a broader range of complex quantum dynamics. In this regard, some recent work even utilized counterdiabatic protocols to simulate quantum critical dynamics on digital quantum hardware~\cite{visuri2025digitized, grabarits2025universal}.
This could enable more sophisticated feature mappings, richer interaction models, and tailored quantum evolutions beyond the constraints of current hardware. Potential candidate platforms include digital superconducting quantum circuits, trapped-ion quantum devices, and even bosonic sampling architectures. Additionally, exploring other forms of analog quantum hardware, such as neutral-atom quantum systems, could provide new opportunities to harness distinctive interaction patterns, further broadening the scope and impact of our method.

\bibliographystyle{unsrt}
\bibliography{reference.bib}

\end{document}